# Polarization Retarder with Cylindrical-Symmetry for Radially Polarized Light

*P.B. Phua[1,2], W.J. Lai[2], Yuan Liang Lim [1]*

[1] *DSO National Laboratories, 20 Science Park Drive, Singapore 118230.*
[2] *Nanyang Technological University, 1 Nanyang Walk, Blk 5 Level 3, Singapore 637616.*

**Abstract:**

We demonstrate a cylindrically-symmetric polarization retarder for generating exact radially polarized light. It involves an uni-axial crystal whose crystallographic optics axis is aligned to its optical axis. This method offers high power-handling capability.

# Polarization Retarder with Cylindrical-Symmetry for Radially Polarized Light

P.B. Phua[1,2], W.J. Lai[2], Yuan Liang Lim[1]
[1] DSO National Laboratories, 20 Science Park Drive, Singapore 118230.
[2] Nanyang Technological University, 1 Nanyang Walk, Blk 5 Level 3, Singapore 637616.

Radially polarized light can be focused tighter [1, 2] for many applications, such as laser machining, laser lithography, optical data storage, high resolution microscopy and etc. It has attracted much attention from various research groups worldwide. What currently lacking is a direct polarization converter that changes an input polarized light into an exact radially polarization profile with high-power laser handling of more than kilowatts.

In this effort, we demonstrated that a polarization retarder with cylindrical-symmetry can generate doughnut-shaped radially polarized beam experimentally. This polarization retarder is an uni-axial crystal whose crystallographic optics axis is aligned to the optical axis of the retarder. An input collimated light of circular polarization is launched into a conical prism (also known as "axicon") and formed a cone of ray. These rays make the same angle, $\theta$, with the crystallographic optics axis of the retarder. Due to this cylindrical symmetry about the crystal's optics axis, all rays experience polarization rotation of the same retardation angle of $\phi = \frac{2\pi}{\lambda}\Delta n(\theta) l(\theta)$. The eigen-axes of this polarization retarder (when projected onto the beam cross-section) also have cylindrical symmetry with the slow (fast) axis in the radial (tangential) direction respectively if the crystal is positive uni-axial. The dashed arrows in Figure 1a illustrate the radial slow axes. If the length of the crystal is chosen so that $\phi = 90°$ (i.e. a quarter-wave), the input circularly polarized light is transformed into an output State-Of-Polarization (SOP) as shown as solid arrows in Figure 1a. Thus, a 45° optical activity quartz rotator can subsequently transform such polarization profile into a radially polarized light. The main advantages of the proposed scheme are 1) it generates an exact continuous radial polarization profile ; 2) since it relies on the cylindrical-symmetry of the uni-axial crystal, there is no discontinuous edge on the optics that leads to diffraction loss [1];  3) it allows high power handling since all optics used can be of high damage thresholds.

In our experiment (see Figure 1b), the laser wavelength used is 532nm and a collimated input circularly light is launched into a cone of rays using a 140° plano-convex fused silica axicon. The uni-axial crystal used is crystalline quartz with thickness of 1.28 mm. Each ray of the cone makes an angle of 6° with respect to the crystal's optics axis. Using a series of lenses, the output beam is captured by the CCD camera. Without a polarizer, the output is a doughnut-shaped light beam as shown in Figure 2 (a). When the polarizer is inserted prior to the camera, two spots are clearly seen, and they rotate with the transmitting axis of the polarizer, as seen in Figure 2 (b)-(e). Hence a radially polarized light has been generated.

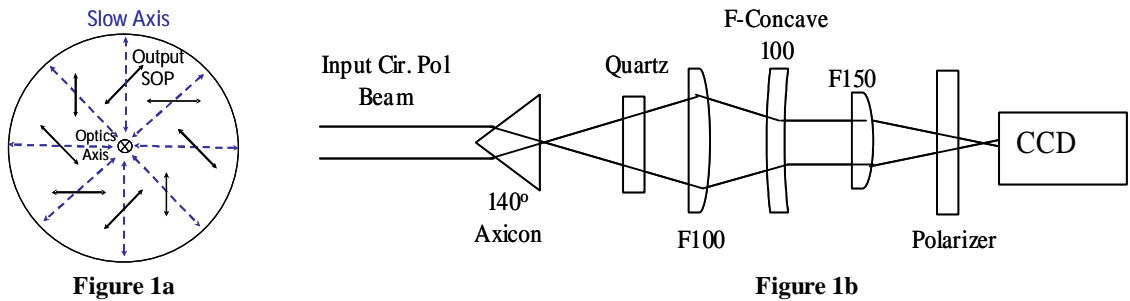

**Figure 1a**     **Figure 1b**

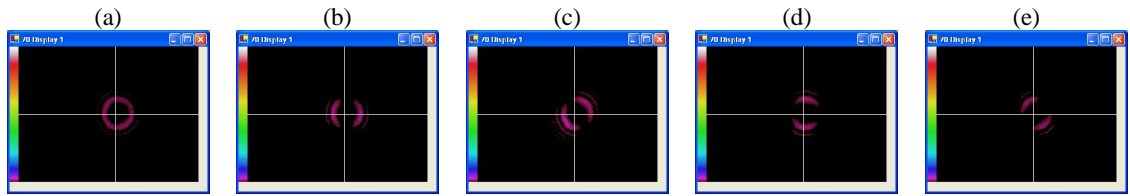

**Figure 2:** With (a) no polarizer and with polarizer rotated to (b) 0°, (c) 45°, (d) 90° and (e) 135°

References:
[1] R. Dorn, S. Quabis and G. Leuchs, Physical Review Letters, Vol. 91, pp. 233901, 2003
[2] I.J. Cooper, M. Roy and C.J.R. Sheppard, Optics Express, Vol. 13, pp. 1066-1071, 2005